\documentclass[twocolumn,superscriptaddress]{revtex4}
\usepackage{mathrsfs}
\usepackage{amssymb,amsbsy,amsmath,latexsym,dsfont,array,layout,graphicx,mathrsfs,color}
\usepackage{multirow}
\usepackage{footnote}
\usepackage{setspace}
\usepackage{subfigure}

\newcommand{\ket}[1]{\left|{#1}\right\rangle}
\newcommand{\bra}[1]{\left\langle{#1}\right|}

\begin{document}

\title{Maximal coin-walker entanglement in a ballistic quantum walk}
\author{Rong Zhang}\altaffiliation{These authors contributed equally to this work.}
\affiliation{National Laboratory of Solid State Microstructure, School of Physics, School of Electronic Science and Engineering, and Collaborative Innovation Center of Advanced Microstructures, Nanjing University, Nanjing, 210093, China}
\affiliation{College of Electronic and Optical Engineering, Nanjing University of Posts
and Telecommunication, Nanjing 210023, China}

\author{Ran Yang}\altaffiliation{These authors contributed equally to this work.}
\affiliation{National Laboratory of Solid State Microstructure, School of Physics, School of Electronic Science and Engineering, and Collaborative Innovation Center of Advanced Microstructures, Nanjing University, Nanjing, 210093, China}

\author{Jian Guo}
\affiliation{National Laboratory of Solid State Microstructure, School of Physics, School of Electronic Science and Engineering, and Collaborative Innovation Center of Advanced Microstructures, Nanjing University, Nanjing, 210093, China}

\author{Chang-Wei Sun}
\affiliation{National Laboratory of Solid State Microstructure, School of Physics, School of Electronic Science and Engineering, and Collaborative Innovation Center of Advanced Microstructures, Nanjing University, Nanjing, 210093, China}

\author{Jia-Chen Duan}
\affiliation{National Laboratory of Solid State Microstructure, School of Physics, School of Electronic Science and Engineering, and Collaborative Innovation Center of Advanced Microstructures, Nanjing University, Nanjing, 210093, China}

\author{Heng Zhou}
\affiliation{School of Information and Communication Engineering, University of Electronic Science and Technology of China, Chengdu 611731, China}

\author{Zhenda Xie}\email{xiezhenda@nju.edu.cn}
\affiliation{National Laboratory of Solid State Microstructure, School of Physics, School of Electronic Science and Engineering, and Collaborative Innovation Center of Advanced Microstructures, Nanjing University, Nanjing, 210093, China}

\author{Ping Xu}\email{pingxu520@nju.edu.cn}
\affiliation{Institute for Quantum Information and State Key Laboratory of High Performance Computing, College of Computing, National University of Defense Technology, Changsha, 410073, China}
\affiliation{National Laboratory of Solid State Microstructure, School of Physics, School of Electronic Science and Engineering, and Collaborative Innovation Center of Advanced Microstructures, Nanjing University, Nanjing, 210093, China}

\author{Yan-Xiao Gong}\email{gongyanxiao@nju.edu.cn}
\affiliation{National Laboratory of Solid State Microstructure, School of Physics, School of Electronic Science and Engineering, and Collaborative Innovation Center of Advanced Microstructures, Nanjing University, Nanjing, 210093, China}

\author{Shi-Ning Zhu}
\affiliation{National Laboratory of Solid State Microstructure, School of Physics, School of Electronic Science and Engineering, and Collaborative Innovation Center of Advanced Microstructures, Nanjing University, Nanjing, 210093, China}


\begin{abstract}
  We report the position-inhomogeneous quantum walk (IQW) can be utilized to produce the maximal high dimensional entanglement while maintaining the quadratic speedup spread of the wave-function. Our calculations show that the maximal coin-walker entanglement can be generated in any odd steps or asymptotically in even steps, and the nearly maximal entanglement can be obtained in even steps after $2$. We implement the IQW by a stable resource-saving time-bin optical network, in which a polarization Sagnac loop is employed to realize the precisely tunable phase shift. Our approach opens up an efficient way for high-dimensional entanglement engineering as well as promotes investigations on the role of coin-walker interactions in QW based applications.
\end{abstract}

\maketitle
\section{INTRODUCTION}
Quantum walk (QW) is the quantum version of classical random walk (CRW)~\cite{ADZ93},
which is considered as a universal computational primitive~\cite{childs09,childs13,YD18}.
The quantum walker can
be in a coherent superposition of many possible position states
resulting in the ballistic transport which is a quadratic speedup
compared with the diffusive spread of CRW~\cite{Kempe03}.
In particular, by adding control to QW dynamics,
the inhomogeneous QW (IQW)
can behave a variety of transport features, including the ballistic behavior, super diffusion,
and localization~\cite{AS11,AJ19,AMBB12,AS10}.
The IQW is also a powerful tool for
investigating interactions and dynamics of quantum particles~\cite{ACH12,CM13},
simulating of complex phenomenons~\cite{TW12,XXue17,sss18,CLP18,XG18},
and engineering of quantum states~\cite{TF19,IK05,GD06,AR10,CT11,SY12,REG13,RG14,QXL18,AA20,AT20,YG19}.

The evolution of a QW is a dynamical interaction process,
resulting in the coin-walker entanglement, or called coin-position entanglement,
which has been studied in a variety of QWs
~\cite{IK05,GD06,AR10,CT11,SY12,REG13,RG14,QXL18,AA20,AT20}.
As spanned in a large number of position states, the entanglement is
in a high-dimensional Hilbert space, and can be used to enhance the test of
quantum fundamentals~\cite{TN10,NS14} and the performance in
quantum communication~\cite{XL02,AA02} and computation~\cite{HN09,EC14,BW11}.
In a particular case that the coin is another degree of freedom of the walker,
it is called the hybrid-entanglement, which is also a valuable resource in quantum information processing
and has broad applications~\cite{OS13,JK08,DT20,KK08,WM03,CZ00,MZ00}.
Besides, the the ballistic spread and quantum entanglement
can be employed to improve the efficiency and performance of the quantum algorithms~\cite{Kempe03,BW11}.
All these applications usually require an efficient way for maximal entanglement generation.
The Hadamard QW, though exhibits the ballistic transport,
namely, the quadratic speedup, cannot reach the maximal coin-walker entanglement~\cite{GD06,AR10}.
In contrast, whereas the maximal entanglement can be created
asymptotically by introducing dynamical disorder~\cite{REG13,QXL18},
or after $10$-step evolution
via an optimal set of coin operations~\cite{AA20,AT20},
these QWs can only behave
sub-ballistic transports with the decreased spread velocity.
Up to now, the simultaneous realization of maximal entanglement and quadratic speedup in the QW has remained a challenge.

In this work, we utilize the position-inhomogeneous quantum walk (IQW)
to produce the maximal entanglement
after any odd steps or large even steps asymptotically,
and the nearly maximal one can be obtained after $2$-step evolution.
Meanwhile, the constructive interference is preserved
leading to the ballistic spread of the wave-function
in the high-dimensional Hilbert space.
Thereby the coin can be correlated to the walker carrying more positions
after the same steps evolution in contrast to the previous schemes with sub-ballistic spread.
Further, we demonstrate our scheme
by implementing the IQW with a high-stable time-bin encoded optical network,
in which a stable polarization Sagnac interferometer
is developed to perform the precisely controlled position-dependent dynamics evolution.
Based on the well-established equivalence of the evolution of coherent light in a linear
optical network and that of a single photon (see p. 106 in \cite{Paulbook}),
here the dimmed pulse laser is employed to simulate the dynamic evolution of single photons.
We simultaneously observe the quadratic speedup transport and maximal entanglement entropy
through the time-multiplexing measurement and the coin state tomography.
Our approach opens up an efficient way to engineer the maximal high-dimensional entanglement
which is applicable in a variety of different physical systems such as atoms\cite{MK09},
trapped ions~\cite{HS09,FZ10} and photons~\cite{SO12,MC18}.

\section{The maximal entanglement with constructive interference}
A quantum walk is defined on a bipartite system
composed of a quantum coin and a walker.
The quantum coin is a two-level system
with $\ket{0}$ ($\ket{1}$) dictating the left (right) moving direction
of the walker for the next step.
The position state of the walker at node $x$ is expressed as
$\ket{x}$ $(x\in \mathbb{Z})$.
The evolution operator of the IQW is defined as
\begin{equation}
\label{eq:U}\hat{U}=\sum_{x}\hat{S}_{x} \left[\hat{C}(x) \otimes \hat{I}\right],
\end{equation}
where $\hat{I}$ is the identity operator on the walker,
and $\hat{C}(x)$ denotes the position-dependent coin operation.
Here the coin operation performed at the position $x=0$ is
the product of the Hadamard operation and a phase shift
\begin{equation}
\label{eq:C0}\hat{C}(0)=\frac{1}{\sqrt{2}}
\begin{pmatrix}
        1 & 1\\
        1 & -1
\end{pmatrix}
\begin{pmatrix}
        e^{i\phi} & 0\\
        0 & e^{i\phi}
\end{pmatrix},
\end{equation}
with $\phi\in[0,2\pi)$.
While at the other positions $x \neq 0$,
$\hat{C}(x)$ is the Hadamard operation.
Thus, this IQW becomes the Hadamard QW (HQW)
when $\phi=0$ is chosen.
The state-dependent shift operation $\hat{S}_{x}$ makes the walker
move to the neighboring left (right) node
determined by the coin state $\ket{0}$ ($\ket{1}$),
which is expressed as
\begin{equation}
\label{eq:Sx}\hat{S}_{x}=\ket{0}\bra{0} \otimes \ket{x-1}\bra{x} + \ket{1}\bra{1} \otimes \ket{x+1}\bra{x}.
\end{equation}
The nonlocal operation $\hat{S}_x$ leads to the coin-walker entanglement.

We consider the initial coin-walker state is a localized state at the original position $x=0$, written as
\begin{align}
\ket{\Psi(0)}=\left[a(0,0)\ket{0}+ b(0,0)\ket{1}\right]\ket{0},
\end{align}
with the complex coefficients satisfying $|a(0,0)|^2+|b(0,0)|^2=1$.
After $t$-step evolution described by $\ket{\Psi(t)}=\hat{U}^{t}\ket{\Psi(0)}$,
the final state can be expressed as
\begin{align}
\label{eq:psit}\ket{\Psi(t)}=\sum_{x}\left[a(x,t)\ket{0}+b(x,t)\ket{1}\right]\ket{x},
\end{align}
with $x=-t,-t+2,\cdots,t-2,t$.
The amplitudes $a(-1,t)$ and $b(1,t)$ can be calculated by
\begin{align}
\label{eq:-1}a(-1,t)=&\frac{1}{\sqrt{2}}e^{i\phi}\left[ a(0,t-1)+b(0,t-1)\right],\\
\label{eq:1}b(1,t)=&\frac{1}{\sqrt{2}}e^{i\phi} \left[a(0,t-1)-b(0,t-1)\right],
\end{align}
and the other amplitudes derived from $x\neq 0$ are
\begin{align}
\label{eq:x-1}a(x-1,t)=&\frac{1}{\sqrt{2}}\left[a(x,t-1)+ b(x,t-1)\right],\\
\label{eq:x+1}b(x+1,t)=&\frac{1}{\sqrt{2}}\left[a(x,t-1)- b(x,t-1)\right].
\end{align}
It is clear that we have $\sum_x |a(x,t)|^2+|b(x,t)|^2=1$ due to the normalized condition.

The evolved coin-walker state remains pure as a result of the unitary evolution together with the pure initial state,
so the entanglement of $\ket{\Psi(t)}$ can be quantified through the von Neumann entropy.
The von Neumann entropy is defined as $E\left(\rho_{\text{c}}\right)=-\text{Tr}(\rho_{\text{c}} \text{log}_2{\rho_{\text{c}}})$
with $1$ and $0$ for the maximally entangled and separable states, respectively,
where $\rho_{\text{c}}(t)$ is the reduced density matrix of coin state obtained by the partial trace over the walker state
$\text{Tr}_{\text{P}} \left[\ket{\Psi(t)}\bra{\Psi(t)}\right]$.
According to the state given by Eq.~(\ref{eq:psit}), we obtain
\begin{equation}
\rho_c=
\begin{pmatrix}
        A(t) & C(t)\\
        C(t)^{*} & B(t)
\end{pmatrix},
\end{equation}
with $A(t)=\sum_x \left|a(x,t)\right|^2$, $B(t)=\sum_x \left|b(x,t)\right|^2$, $C(t)=\sum_x a(x,t)b(x,t)^*$.
Then the von Neumann entropy is calculated by
\begin{equation}
\label{eq:Elambda}E\left(\rho_{\text{c}}\right)=-\lambda_1 \text{log}_{2}{\lambda_1}-\lambda_2 \text{log}_{2}{\lambda_2},
\end{equation}
where $\lambda_{1,2}$ are two eigenvalues of $\rho_{\text{c}}$ expressed as
\begin{equation}
\lambda_{1,2}=\frac{1}{2}\left\{1\pm\sqrt{1+4\left[|C(t)|^2-A(t)B(t)\right]}\right\}.
\end{equation}
An necessary condition of obtaining the maximal entanglement
is that the two diagonal matrix elements of $\rho_{\text{c}}$ equal $1/2$~\cite{AR10},
and hence here we choose a typical balanced state as the initial state which is given by
\begin{equation}
\ket{\Psi(0)}=\frac{1}{\sqrt{2}}(\ket{0}+ e^{i\theta}\ket{1})\ket{0},
\end{equation}
with $\theta \in [0,2\pi)$. Therefore, we have $A(t)=B(t)=1/2$, and $\lambda_{1,2}=1/2\pm|C(t)|$.
So the entanglement entropy $E$ depends on the initial state, coin operations.
and evolution steps.

\begin{figure}
    \includegraphics[width=4.2cm]{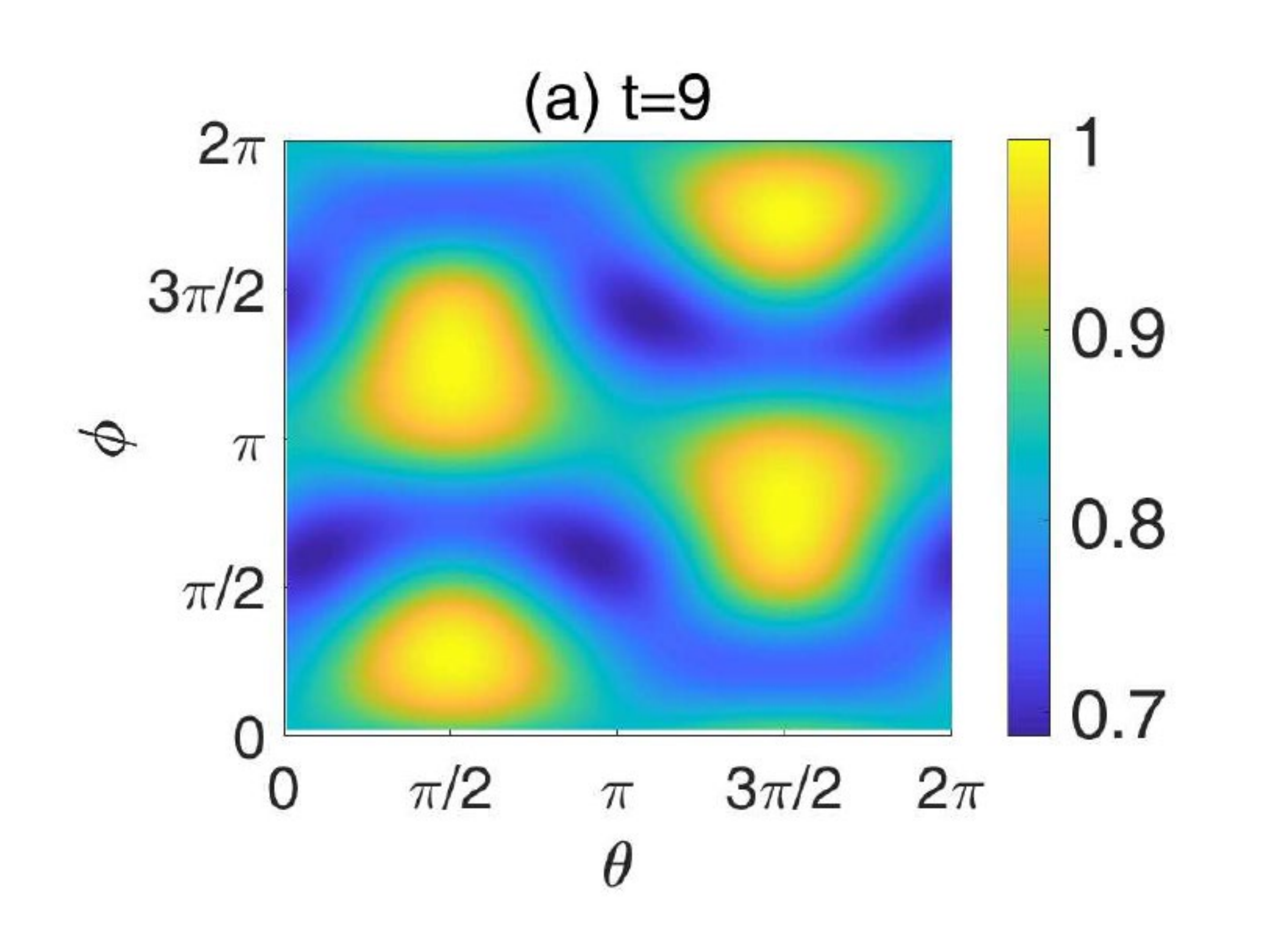}
    \includegraphics[width=4.2cm]{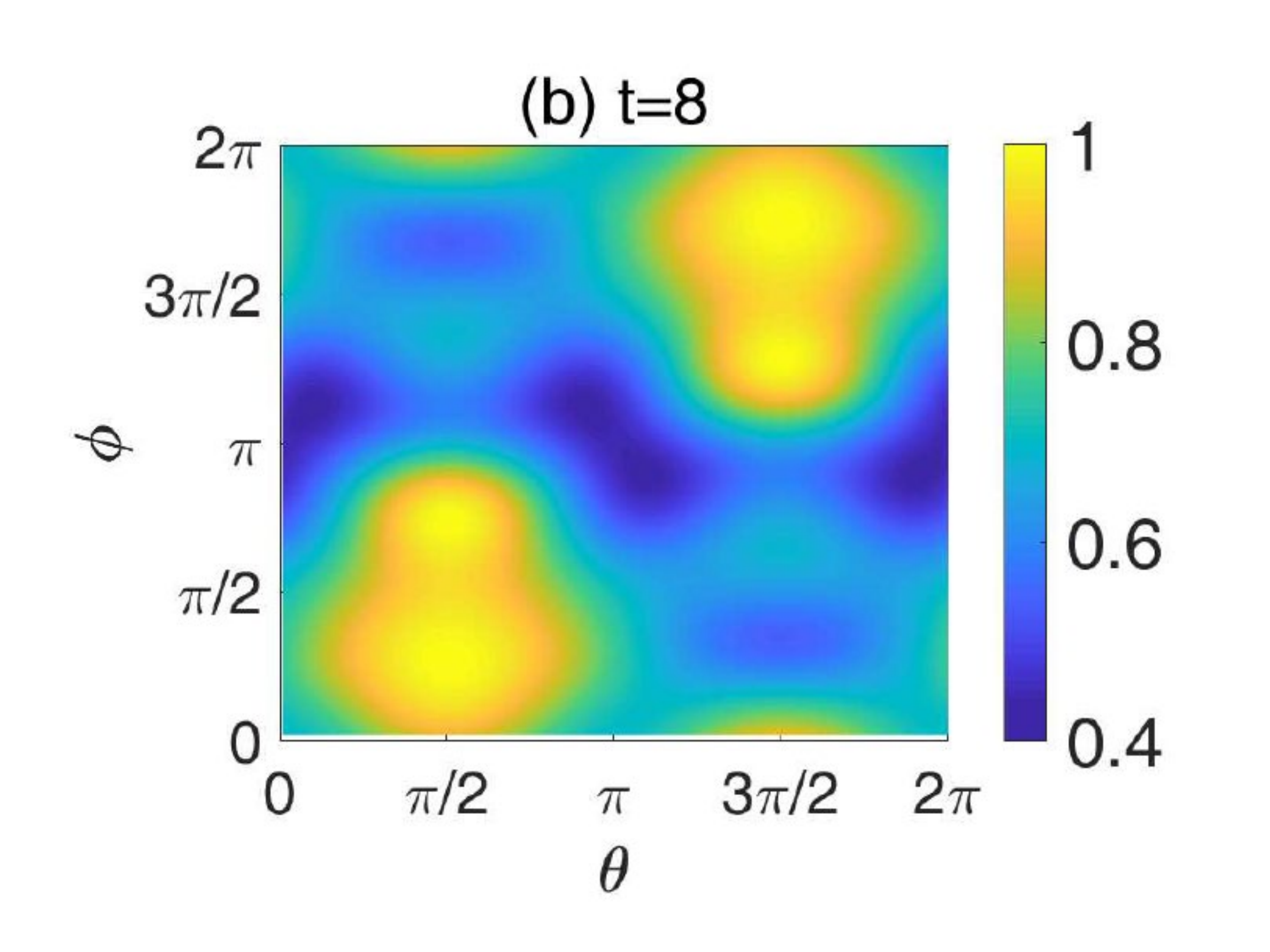}
    \includegraphics[width=8cm]{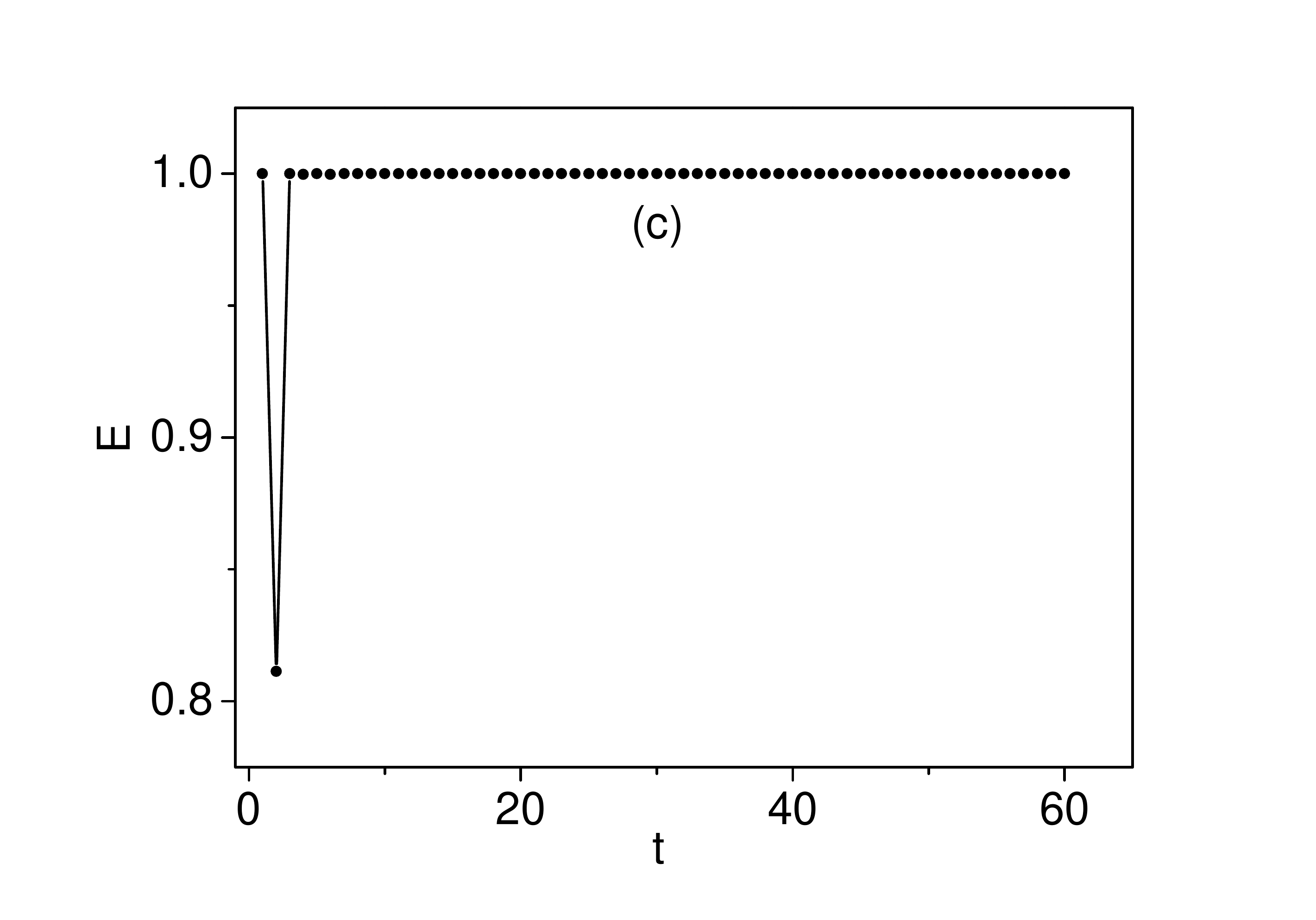}
\caption{(Color online.)
     The von Neumann entropy of the coin-walker entanglement $E$ as a function of the initial state parameter $\theta$ and the coin operation parameter $\phi$ of the inhomogeneous quantum walk for the steps of (a) $t=9$ and (b) $t=8$. (c) The von Neumann entropy as a function of evolution step $t$ with $\theta=\pi/2$ and $\phi=\pi/4$.}
\end{figure}

Then we investigate the von Neumann entropy
as a function of $\theta$ and $\phi$ for any step
with the amplitude recursion formulas given by Eqs.~(\ref{eq:-1})-(\ref{eq:x+1}).
Taking $t=9$ and $t=8$ for example, the calculation results are
shown in Fig.~1 (a) and (b), respectively.
We can see that the maximal value of the von Neumann entropy can be obtained
by choosing $\phi=\pi/4$, $\theta=\pi/2$, or $\phi=7\pi/4$, $\theta=3\pi/2$.
Actually, the two conditions can maximize the von Neumann entropy for all the steps except step $2$.
Taking either condition, here choosing the former one, we plot the entanglement entropy
for $60$ steps in Fig.~1 (c), and list the von Neumann entropy values for $11$ steps in Tab.~1 (Column $E_{\text{IQW}}^{\text{theo}}$).
It is shown that for any odd steps, maximal entanglement $E=1$ can be obtained,
while for the even steps, the maximal entanglement is achieved asymptotically and
nearly maximal entanglement with $E>0.99$ has been achieved since step $2$.
For comparison, we also list the von Neumann entropy values for the HQW in Tab.~1
(Column $E_{\text{HQW}}^{\text{theo}}$), which are all around $0.87$.

\begin{figure}[t]
    \includegraphics[width=0.5\textwidth]{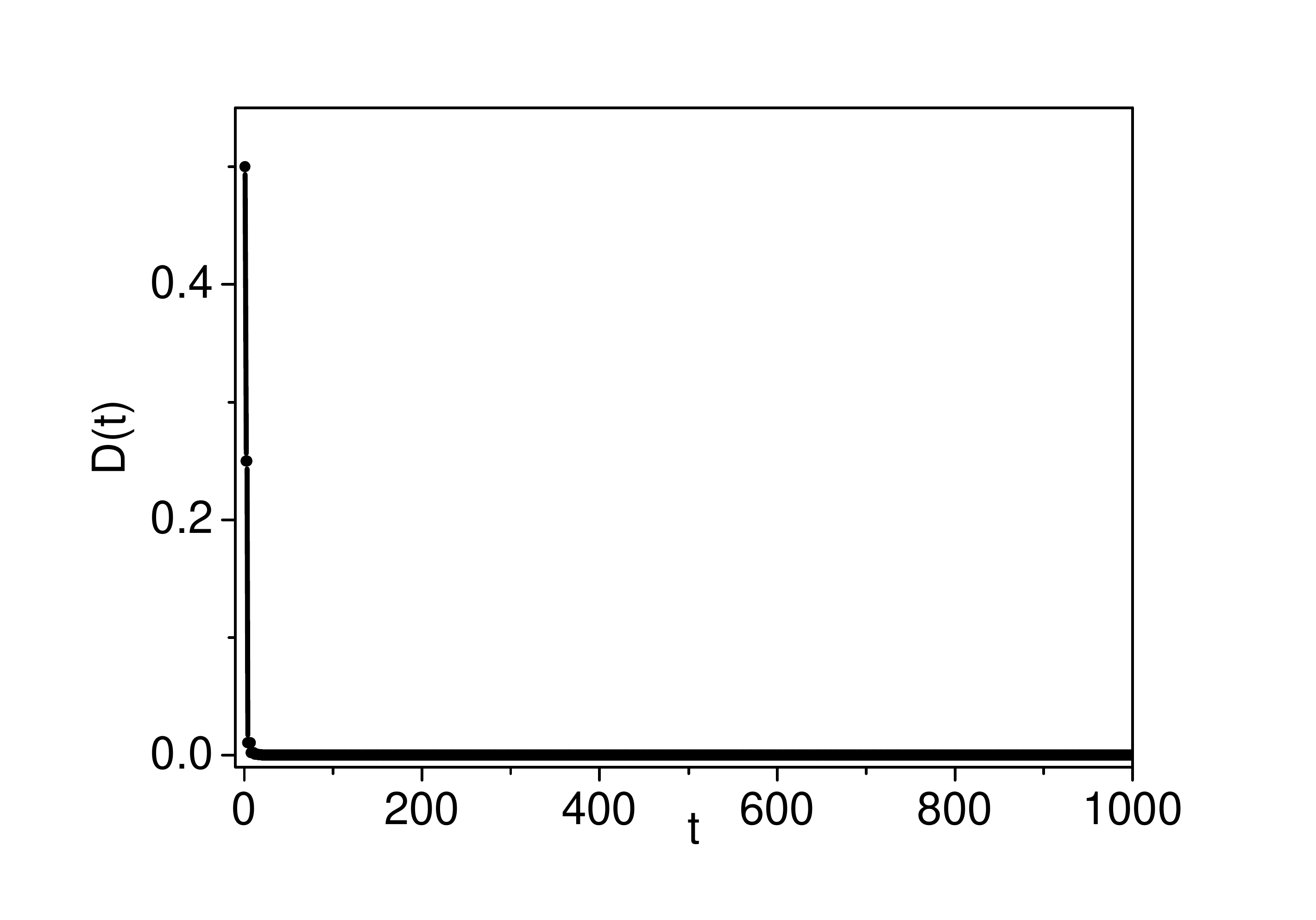}
    \caption{Trace distance between two neighboring coin states as a function of evolution step $t$ with $\theta=\pi/2$ and $\phi=\pi/4$.}
\end{figure}

To further investigate the asymptotic limit of entanglement in even steps,
we calculate the trace distance between two coin states in adjacent steps by
\begin{align}
  D(t)=\frac{1}{2}\text{Tr}\left[\left|\rho_{\text{c}}(t)-\rho_{\text{c}}(t-1)\right|\right],
\end{align}
where $\left|M\right|=\sqrt{M^{\dag}M}$ for a matrix $M$.
For the qubit state $\rho_{\text{c}}(t)$, the trace distance $D(t)$ is equal to
the Ky Fan $1$-norm, i.e., the largest singular value of $\rho_{\text{c}}(t)-\rho_{\text{c}}(t-1)$~\cite{MI00}.
In Fig.~2, we plot the calculation results of $D(t)$ in up to $1000$ steps,
and then get a nonlinear fitting function of $D(t)\sim t^{-1.90}$.
Consequently, we have $\text{lim}_{t \to \infty}\left[\rho_{\text{c}}(t+1)-\rho_{\text{c}}(t)\right]=0$.
Namely, $\rho_{\text{c}}(t+1)=\rho_{\text{c}}(t)$ works in the large step regime. As for any odd steps maximal entanglement is obtained, and hence maximal entanglement is achieved in the large even step regime as well.

The signature of quantum effects is also imaged in
the walker's spread velocity quantified by the position variance
$\nu(t)=\langle x^2\rangle-\langle x\rangle^2$.
We show the calculation results of the position variance for the above IQW up to $1000$ steps
in Fig.~3 (the black solid curve), and for comparison the position variance values
of the HQW (the red dashed curve) and CRW (the blue dotted line) are given as well.
One can see that the IQW and HQW can both exhibit the ballistic transport due to the maintained constructive interference,
in contrast to the diffusive spread of CRW. Moreover, the IQW spreads even faster than the HQW.

\begin{figure}[t]
\centering
\includegraphics[width=0.48\textwidth]{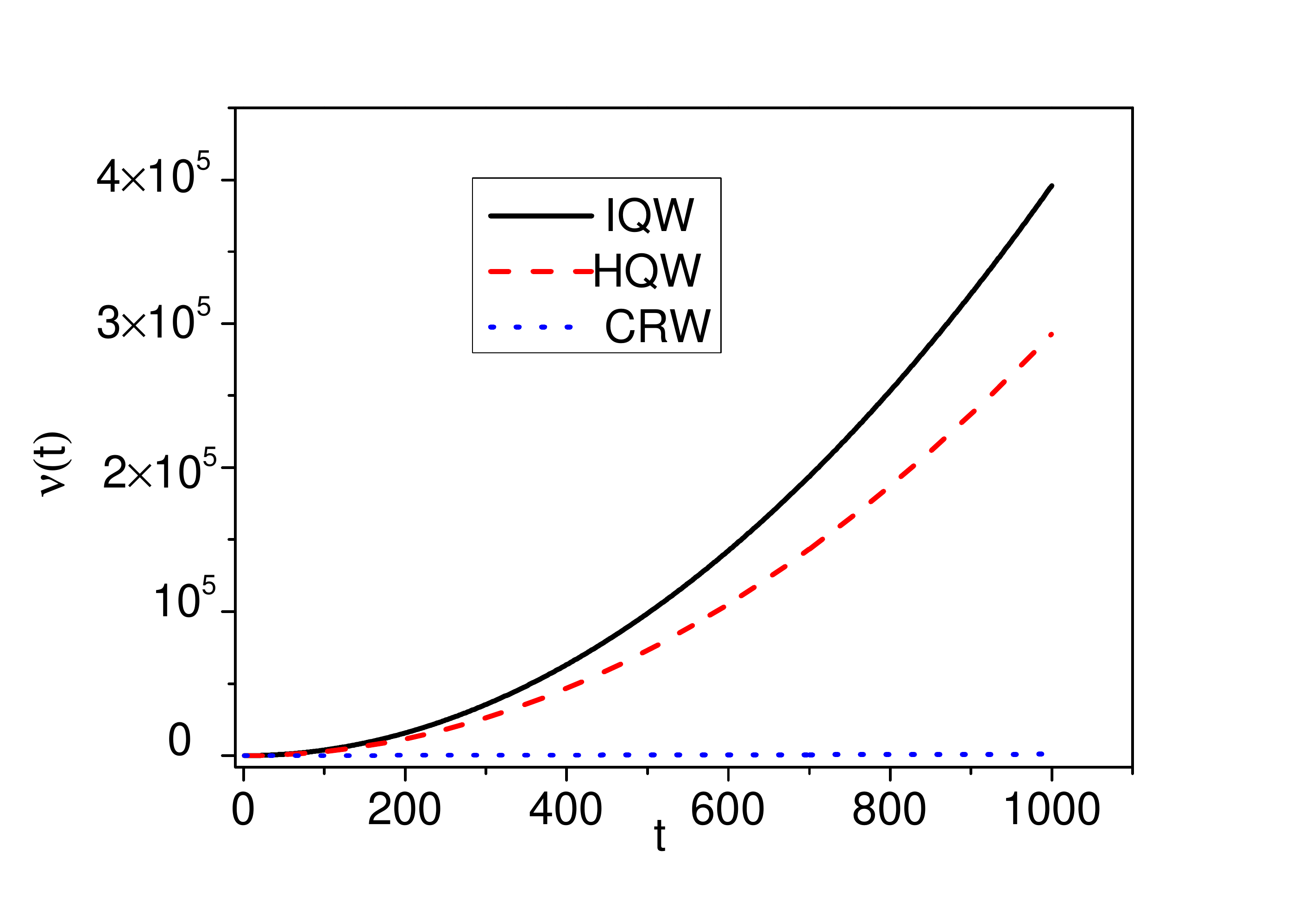}
\caption{Position variance as a function of evolution step $t$ with $\theta=\pi/2$ and $\phi=\pi/4$.}
\end{figure}

\section{Experimental demonstrations}
We realize the IQW defined in Eqs.~(\ref{eq:U})-(\ref{eq:Sx})
by the resource-saving time-bin encoded optical network as sketched in Fig.~4. Our setup consists of three parts, the initial state preparation (the green part),
the evolution (the blue and pink parts),
and the state measurement (the yellow part).
The horizontal and vertical polarization states $\ket{\text{H}}$ and $\ket{\text{V}}$
represent the coin states $\ket{0}$ and $\ket{1}$, respectively.
The position states are encoded in discrete time bins
formed by distinct path lengths.
We use an electro-modulated picosecond diode laser (Hamamatsu Photonics, PLP-10),
with a pulse width of $70$~ps, a repetition rate of $1$~MHz, and a central wavelength of $1550$~nm.
The laser output first passes through a polarization beam splitter (PBS0),
a half-wave plate (HWP0),  PBS1 and the light intensity can be controlled
by tuning the optical axis angle of the HWP0.
Then the light is coupled into the evolution of quantum walk via
the $1/99$ port of a $1:99$ single-mode fiber beam splitter (BS1)
after passing through HWP1 and the quarter-wave plate (QWP1), where the two wave-plates can
prepare an arbitrarily initial polarization of the light output from BS1.

\begin{figure}[t]
 \includegraphics[width=0.48\textwidth]{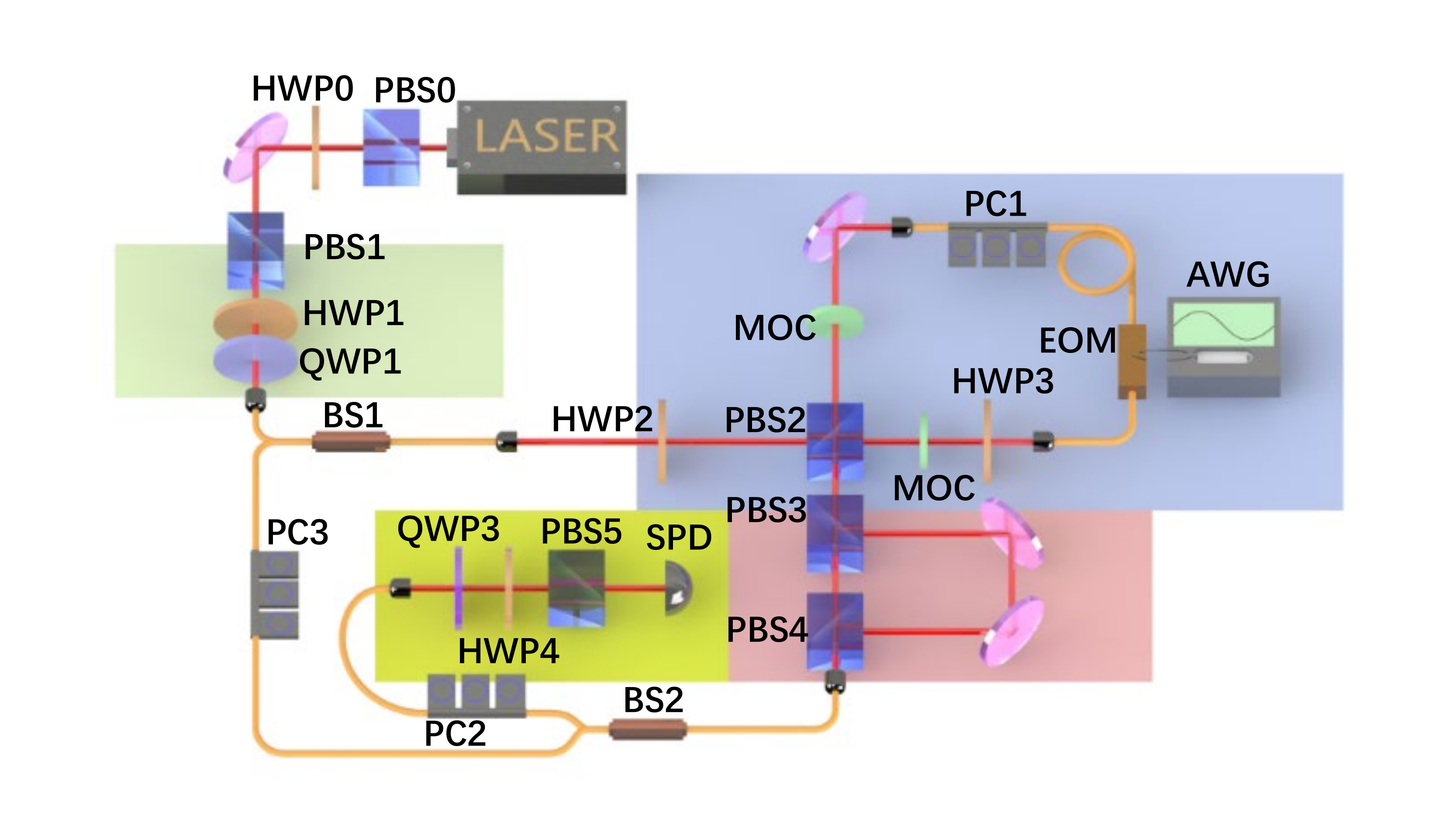}
  \caption{Experimental setup of the ordered inhomogeneous quantum walk.
    PBS: polarization beam splitter; HWP: half-wave plate; QWP: quarter-wave plate;
    BS: single-mode fiber beam splitter; PC: polarization controller;
    MOC: magneto-optic crystal; EOM: electro-optic phase modulator;
    AWG: arbitrary wave generator; SPD: single-photon detector.
    }\label{Fig:setup}
\end{figure}

The evolution of a single-step IQW is realized by a round trip through the optical network
including the variable coin operations (the blue part)
and state-dependent shift operator (the pink part).
The position-dependent coin operations given in Eq.~(\ref{eq:C0})
are realized through HWP2 followed by a polarization Sagnac loop.
Explicitly, the optical axis of HWP2 is oriented at $22.5^{\circ}$ to perform the Hadamard operation.
Because the phase modulations generated directly by the electro-optic
phase modulator (EOM) (Eospace) depend on $H$ and $V$ polarizations with the restriction $\phi_V/\phi_H=3.5$,
the Sagnac loop in the the blue part of Fig.~\ref{Fig:setup} is employed to perform two respective
fast-tunable phase shifts $\phi_V$, $\phi_H$ on the H and V polarization of the input light in a stable way.
Firstly, a beam of light is input from one port of PBS2, and the H and V polarization parts are separated
into counter-clockwise and clockwise paths, respectively.
Secondly, the input port of clockwise light is fused with a single-mode fiber to introduce a $39.1$-ns delay of the
arrival time at the EOM compared with the counter-clockwise light,
so that the phases on H and V polarization can be  precisely controlled
by the voltages on the EOM separately.
Both ports of the EOM are connected by polarization-maintaining (PM) fibers in slow axis aligned.
A magneto-optic crystal (MOC) is placed in each path to make the polarization rotate $45^{\circ}$ counter-clockwise.
The optical axis of HWP3 is tuned to make the diagonal polarization coupling with the slow axis of the PM fiber.
Likewise, the polarization controller PC1 is adjusted to couple the anti-diagonal polarization with the slow axis.
Finally, the EOM was triggered by a programmable electric pulse output from the arbitrary wave generator (AWG) (Tektronix, AWG7012),
and the light in each path returns its original polarization with an individual modulated phase
at the output of the loop.
Here the required phases on H and V polarization are the same $\phi_V=\phi_H=\pi/4$ as shown in Eq.~(\ref{eq:C0}).

The shift operation $\hat{S}_x$ given in Eq.~(\ref{eq:Sx}) is implemented
by an unbalanced polarization Mach-Zehnder interferometer
formed by PBS3 and PBS4,
where the V polarization
is delayed $\Delta t=2.3$~ns compared with the H polarization.
The temporal difference $\Delta t$
corresponds to a step in two positions $x\pm 1$.
Finally, about one percent of photons are coupled out of the network
to the measurement setup
and the majority go to the next round of
the network via a $99:1$ beam splitter BS2,
with a one-loop time of $T=72.9$~ns.
The one-loop loss is measured to be $3.6$~dB.
The measurement setup consists of a single-photon detector (SPD) (Quantique id201) and
a combination of QWP3, HWP4, and PBS5, which can realize any state projection measurement~\cite{DW01,FS15}.
The SPD and the laser are both triggered by the electrical pulses from the AWG. Therefore,
the state evolution of each step can be accurately monitored in our setup.
All experimental results are averaged over more than $10^4$ detected events.

The initial coin-walker state at the original position is prepared as
$\ket{\Psi_{\text{1}}(0)}=(\ket{\text{H}}+ i\ket{\text{V}})\ket{0}/\sqrt{2}$.
As shown in Eqs.~(\ref{eq:U})-(\ref{eq:Sx}), we realize the HQW and IQW
by choosing $\phi=0$, and $\pi/4$, respectively.
For the HQW, the EOM is powered off.
To implement the required IQW, the phase shift $\phi_V=\phi_H=\pi/4$ on both H and V polarizations
are performed at the position $x=0$ of even steps $2m$ corresponding with the time bins of $t_0=m(2T+\Delta t)$ $(m\in \mathbb{Z})$.
So a programmable electrical pules from the AWG with a pulse width of $1$~ns and a pulse height of $1.8$~V
at both time bins $t_0$ and $t_0+39.1$~ns is applied on the EOM
when the H polarization pulse and the V one of the position $x=0$ arrive.

\begin{figure}[t]
    \includegraphics[width=0.48\textwidth]{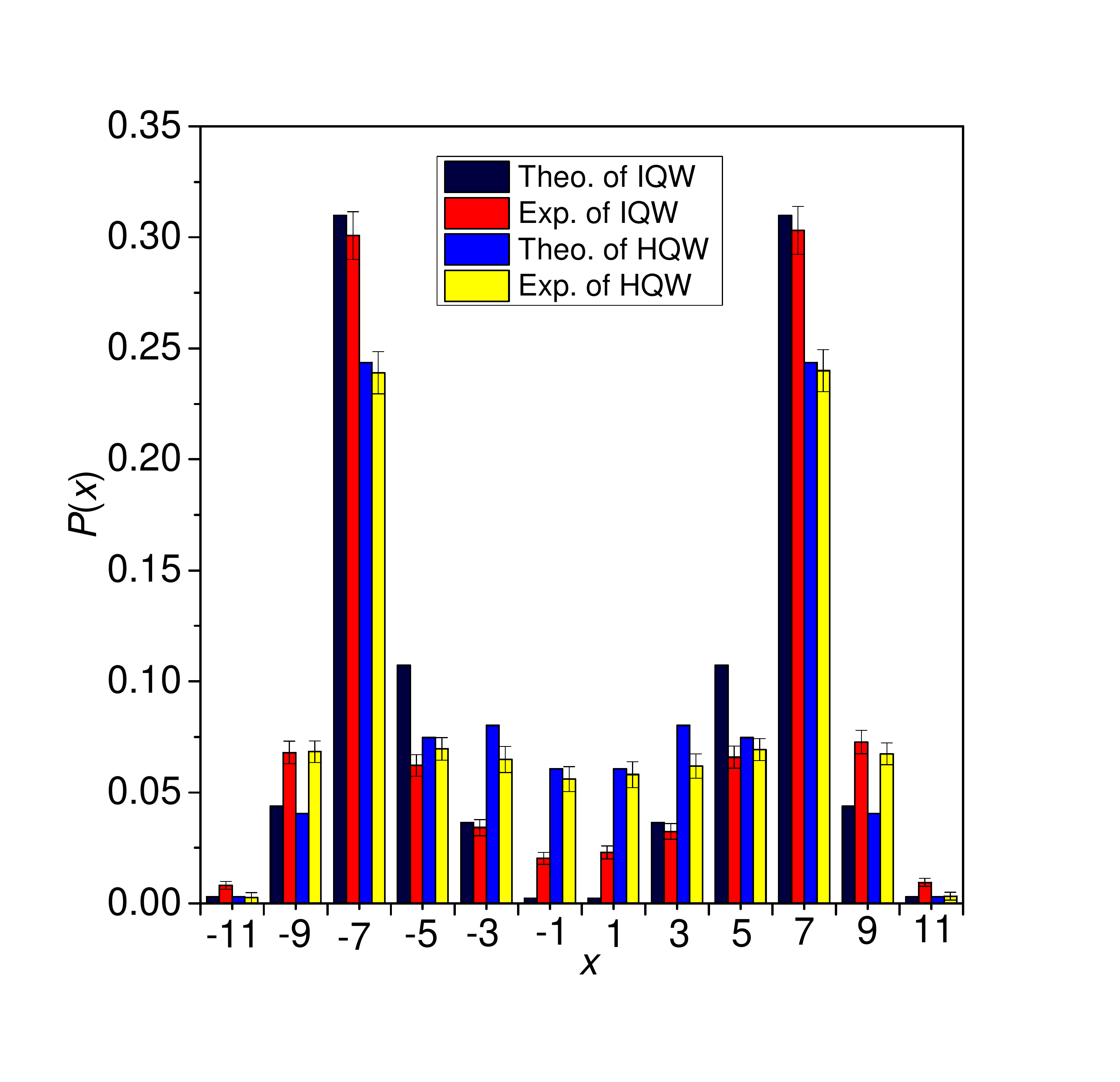}
    \caption{Experimental and theoretical position distributions $P(x)$ of the inhomogeneous quantum walk (IQW)
    and the Hadamard quantum walk (HQW) in step $11$.
    Error bars are simulated from the statistical errors.
    }\label{Fig:distribution}
\end{figure}

After $t$-step evolution, the initial light pulses with the time bin $t=0$ are distributed in $t+1$ time windows
corresponding with the positions $x=-t, -t+2, \cdots, t$. We apply projecting measurement on the polarization bases
$\{\ket{H},\ket{V},\ket{H}+\ket{V},\ket{H}-i\ket{V}\}$ for each time bin.
So we obtain the position distribution probabilities
by adding up the the counts of projection measurement
on state $\ket{\text{H}}$ and $\ket{\text{V}}$ in each position, and then normalizing them with the sum of all the positions.
The experimental results of IQW (red bars) and HQW (yellow bars) in step $11$ are shown in Fig.~5,
with the black and blue bars representing the corresponding theoretical probabilities, respectively.
The position distributions for both HQW and IQW clearly show the non-Gaussian behaviors of the completely coherent quantum walks.
To compare the experimental ($P_{\text{exp}}$) and theoretical ($P_{\text{theo}}$) distributions, we employ the similarity defined by $S = \sum_x \sqrt{P_{\text{exp}}P_{\text{theo}}}$,
which ranges from $0$ for the complete mismatch
to $1$ for the perfect concordance.
The similarities for HQW and IQW are calculated as $S_{\text{H}} = 0.9805\pm 0.0002$, and $S_{\text{I}} = 0.9673\pm 0.0002$, respectively.
Therefore, nearly perfect coherence in our setup
is confirmed by the non-Gaussian distributions after $11$-step evolution.

\begin{table}[b]
\centering
\caption{Experimental results and theoretical values of the
von Neumann entropy of the inhomogeneous quantum walk (IQW)
and the Hadamard quantum walk (HQW). Errors are estimated from the statistical errors.}
\label{Tab:01}
\setlength{\tabcolsep}{7pt}
\renewcommand\arraystretch{1.3}
\begin{tabular*}{8.6cm}{ccccc}\hline\hline
step & $E_{\text{IQW}}^{\text{theo}}$ & $E_{\text{IQW}}^{\text{exp}}$ & $E_{\text{HQW}}^{\text{theo}}$ & $E_{\text{HQW}}^{\text{exp}}$  \\
\hline
$1$  & $1$         &    $0.99\pm0.01$    &  $1$          &  $0.99\pm0.01$ \\
$2$  & $0.81128$   &    $0.86\pm0.01$    &  $0.811$      &  $0.81\pm0.01$ \\
$3$  & $1$         &    $0.99\pm0.01$    &  $0.811$      &  $0.81\pm0.01$  \\
$4$  & $0.99967$   &    $0.95\pm0.02$    &  $0.896$      &  $0.88\pm0.02$  \\
$5$  & $1$         &    $0.95\pm0.02$    &  $0.896$      &  $0.88\pm0.02$  \\
$6$  & $0.99967$   &    $0.99\pm0.02$    &  $0.857$      &  $0.83\pm0.02$  \\
$7$  & $1$         &    $0.99\pm0.02$    &  $0.857$      &  $0.86\pm0.02$  \\
$8$  & $0.99999$   &    $0.97\pm0.03$    &  $0.882$      &  $0.90\pm0.02$  \\
$9$  & $1$         &    $0.99\pm0.04$    &  $0.882$      &  $0.89\pm0.03$  \\
$10$ & $0.99999$   &    $0.99\pm0.05$    &  $0.865$      &  $0.88\pm0.04$  \\
$11$ & $1$         &    $0.99\pm0.06$    &  $0.865$      &  $0.85\pm0.05$  \\
\hline
\end{tabular*}
\end{table}

Based on the experimental results of the position distributions of the two types of QWs, we calculated
the position variance $\nu(t)=\sum_{x}P(x,t)\left|x-\mu(t)\right|^2$ with $\mu(t)=\sum_{x}P(x,t) x$
in order to quantify the spread of wave-function.
The variance of the HQW (blue triangles) and IQW (red dots)
in up to $11$ steps are shown in Fig.~6, with the corresponding theoretical results plotted as solid curves.
For comparison, the variance of CRW is also plotted with the black solid curve. We can see that in contrast to the diffusive spread in CRW,
the IQW as well as the HQW can behave the quadratic speedup spread,
and moreover, the IQW spreads even faster than the HQW.
Taking step $11$ as an example, the position variances of the IQW, HQW and CRW are $\nu_{\text{I}}(11)=46.91\pm 2.99$, $\nu_{\text{H}}(11)=37.46\pm 2.76$, and $\nu_{\text{C}}(11)=11$, respectively.

\begin{figure}[t]
        \includegraphics[width=0.5\textwidth]{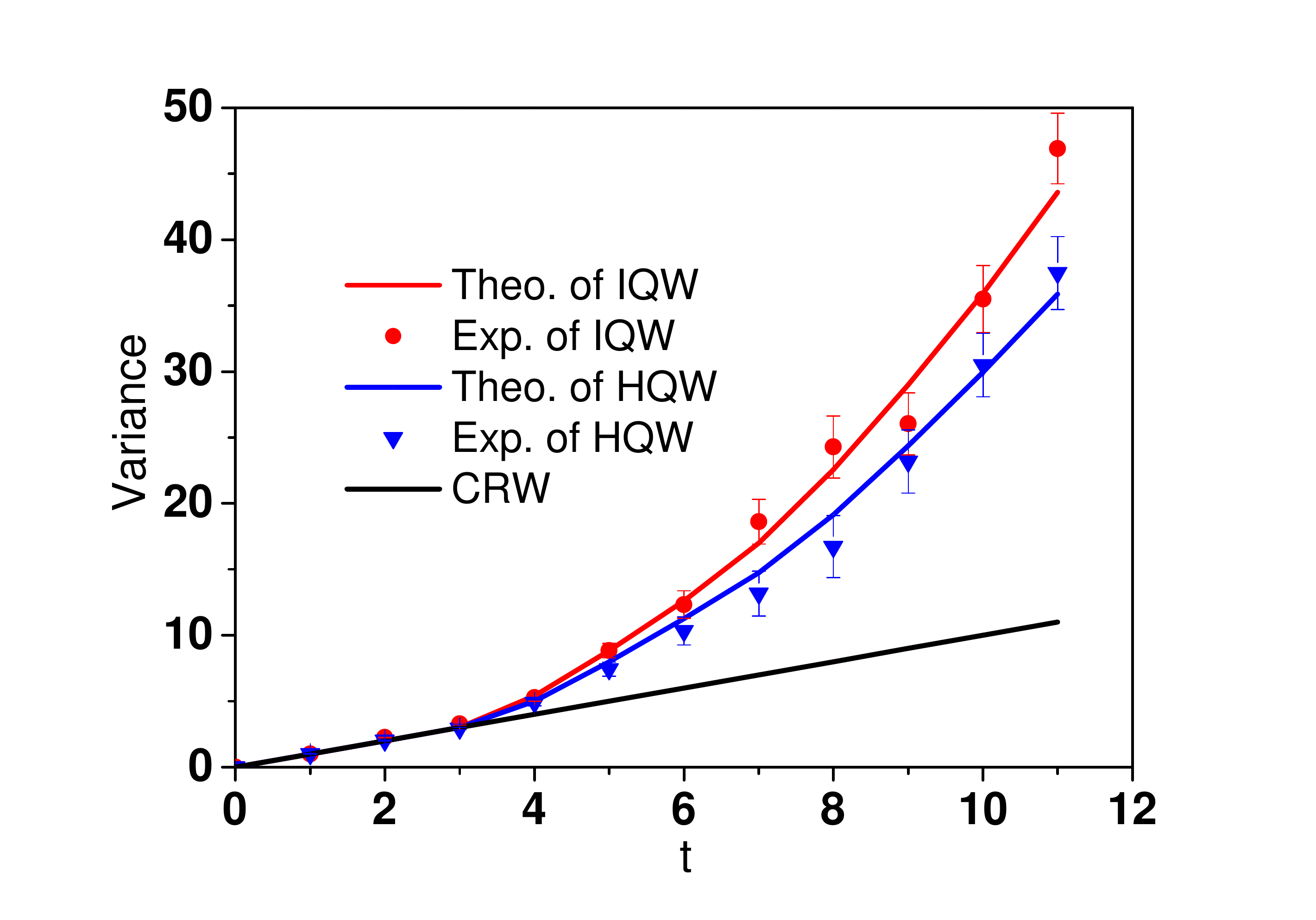}
        \caption{Experimental position variance against the evolution step $t$ for the inhomogeneous quantum walk (IQW), the Hadamard quantum walk (HQW) in up to 11 steps, with the corresponding theoretical solid curves. The black solid curve represents the theoretical position variance of the classical random walk (CRW). Error bars are simulated from the Poissonian statistics.
    }
\end{figure}

The above experimental results, including both the non-Gaussian distribution and quadratic speedup,
have demonstrated that the coherence are maintained perfectly and further confirm that
the coin-walker are in the coherent superposition state during the whole evolution~\cite{BCA03}.
So the von Neumman entropy can be used to quantify the coin-walker entanglement.
The state tomography on the reduced coin density matrix can be obtained
by adding up the counts of all $t+1$ positions for each bases polarization state
$\ket{H}$, $\ket{V}$, $\ket{H}+\ket{V}$ and $\ket{H}-i\ket{V}$.
And we reconstruct all the coin density matrices from step $1$ to $11$ of the HQW and IQW, respectively.
The fidelities, calculated by
$F=\left(\text{Tr}\sqrt{\sqrt{\rho_{\text{c}}^{\text{exp}}}\rho_{\text{c}}^{\text{theo}}\sqrt{\rho_{\text{c}}^{\text{exp}}}}\right)^2$,
are all higher than $F_{\text{H}} =0.967\pm 0.008$ and
$F_{\text{I}} =0.952\pm 0.009$ for HQW and IQW, respectively.
The calculated values of the von Neumman entropy from the experimental results are
listed in Tab.~1, which all agree well with the corresponding theoretical values.
Therefore, it is experimentally convinced that the maximal coin-walker entanglement can be produced
in the ballistic quantum walk.

\section{Conclusion}
We have proposed and experimentally demonstrated that the position-inhomogeneous quantum walk (IQW) can generate maximal coin-walker entanglement as well as behave the quadratic speedup by employing position-dependent coin operations. Our theoretical analysis shows that maximal coin-walker entanglement can be created in any odd steps or asymptotically in even steps, and approximately maximal entanglement can be generated in even steps after $2$ as well. We have implemented the IQW through a stable time-bin-walking optical network, with a fast-tunable polarization Sagnac loop developed for implementing the coin operations.
We observes high-fidelity evolutions in up to $11$ steps and
both the maximal entanglement and quadratic speedup in the IQW have been confirmed experimentally.
Our approach opens up a way for efficient creation of high-dimensional (hybrid) entanglement, which have applications in a variety of quantum technologies. Our investigations can be extended to high-dimensional QW with broader applications. Our work can also prompt the applications such as quantum computation based on the quantum walk with respective to the role of coin-walker interactions.

\section{ACKNOWLEDGEMENTS}
This work was supported by National Key R\&D Program of China (No.
2019YFA0705000 and 2019YFA0308700), the Key R\&D Program of Guangdong Province (Grant No.
2018B030329001), National Natural Science Foundation of China (Grants No. 51890861, No. 11547031,
No. 11705096  No. 11674169, No. 11627810, No. 61705033, No. 11690031, and  No. 11974178).

\section{REFERENCES}


\end{document}